\documentstyle[12pt,psfig,aaspp]{article}
\def\beq{\begin{equation}}
\def\eeq{\end{equation}}
\def\vg{\vec g}
\def\vu{\vec u}
\def\vr{\vec r}
\def\vv{\vec v}
\def\va{\vec a}
\def\grad{\vec\nabla}
\def\div{\vec\nabla\cdot}
\def\a0{$a_0$}
\begin{document}
\title{MOND--a pedagogical review\footnote{Presented at the XXV
 International School of Theoretical Physics ``Particles and
 Astrophysics--Standard Models and Beyond'', Ustro\'n, Poland, 
September 10-16, 2001}}
\author{Mordehai Milgrom }
\affil{ Department of Condensed Matter Physics, Weizmann
Institute, Rehovot, Israel}

\begin{abstract}
An account is given of the development, and the status, of the
modified dynamics (MOND)--a proposed alternative to dark matter,
which posits a breakdown of Newtonian dynamics in the limit of
small accelerations.

\end{abstract}

\section{introduction}
\par
The evidence for dark matter is only indirect. What the evidence
points to directly is a mass discrepancy in galaxies and other
galactic systems: When we count the mass of baryonic matter in
such systems--in stars, neutral and high-T gas, etc.--the total
sum does not provide enough gravity to explain the observed
accelerations in such systems within standard physics. If we
adhere to standard dynamics, the need for dark matter is the only
solution we can conceive. It is, however, possible that the laws
of dynamics, proven in the laboratory and the solar system, cannot
be simply applied in the realm of the galaxies. An appropriate
modification of the laws of dynamics for parameters that are
pertinent to these, might obviate the need for dark matter
altogether, if it produces the observed accelerations with only
the observed baryonic mass distribution.
\par
But, exactly which system attribute makes the difference? Galactic
systems have masses, sizes, and angular momenta that are many
orders of magnitude larger than those in the solar system. The
large distances involved is a natural culprit. Indeed, there were
attempts to modify the distance dependence of gravity: the
gravitational force is still taken as proportional to the two
masses involved but the decline at large distances is not as
strong as in the $r^{-2}$ law. Such a modification cannot,
however, explain away dark matter. If the modified law is to
produce asymptotically flat rotation curves of disc galaxies, as
observed, it automatically predicts the wrong form of the mass
velocity relation: it gives $M\propto V^2$, instead of $M\propto
V^\alpha$, with $\alpha\sim 4$, as required by the observed
Tully-Fisher relation (\cite{milgal}). In even more blatant
conflict with observations, such modifications predict that the
mass discrepancy should increase systematically with system size.
In contrast, dwarf spheroidal galaxies, among the smallest in the
galactic menagerie, show very large mass discrepancies, much
larger then some large galaxies. And, the much larger galaxy
clusters show only moderate mass discrepancies. A semi-schematic
depiction of the systematics of the mass discrepancy with distance
can be seen in Fig 1 in \cite{mildark98}, where it is obvious that
the observed mass discrepancy does not increase systematically
with size.
\par
In the early 1980s I proposed a modified-dynamics based on the
acceleration as the relevant system parameter, based on the fact
that typical accelerations in galactic systems are many orders of
magnitude smaller than those encountered in the solar system.
Since then, a handful of us have been working on the development
of this scheme, which has involved devising more refined theories,
elaborating the observational consequences, and testing them
against the data.

\section{The modified dynamics}
\par
This modified dynamics, MOND, introduces a constant with the
dimensions of an acceleration, \a0, and posits that standard
Newtonian dynamics is a good approximation only for accelerations
that are much larger than \a0. The exact behavior in the opposite
limit is described by the specific underlying theory, to be
described below. However, the basic point of MOND, from which
follow most of the main predictions, can be simply put as follows:
 a test particle at a distance $r$ from a large mass $M$ is
 subject to the acceleration $a$ given by
 \beq a^2/a_0=MGr^{-2}, \label{basic} \eeq
  when
 $a\ll a_0$, instead of the standard expression $a=MGr^{-2}$,
 which holds when $a\gg a_0$.
 The two expressions may be interpolated to give the heuristic
 relation
 \beq \mu(a/a_0)a=MGr^{-2}=a_N, \label{mond} \eeq
 where $a_N$ is the Newtonian expression for the acceleration,
 and the interpolating function $\mu(x)$ satisfies
 $\mu(x)\approx 1$ when $x\gg 1$, and $\mu(x)\approx x$ when $x\ll
 1$. This expression, while lacking from the formal point of view,
 is very transparent, and captures the essence of MOND. I shall
 describe below more presentable theories based on this basic
 relation, but these are still phenomenological theories into
 which the form of $\mu(x)$ has to be put in by hand. It will
 hopefully follow one day from a more basic
 underlying theory for MOND, which we still lack. Most of the
 implications of MOND do not depend strongly on
the exact form of $\mu$. Much of the phenomenology pertinent to
the mass discrepancy in galactic systems occurs in the deep-MOND
regime ($a\ll a_0$), anyway, where we know that $\mu(x)\approx x$.

\section{MOND phenomenology}
\par
One immediate result of eqs.(\ref{basic})(\ref{mond}) is that at a
large radius around a mass $M$, the orbital speed on a circular
orbit becomes independent of radius. This indeed was a guiding
principle in the construction of MOND, which took asymptotic
flatness of galaxy rotation curves as an axiom (even though at the
time it was not clear how definite, and how universal, this is).
Second, this asymptotic rotational speed depends only on the total
mass $M$ via $V^4=MGa_0$. This, according to MOND, is the fact
underlying the observed Tully-Fisher-type relations, by which the
typical (mean) rotational velocity, $V$, in a disc galaxy is
strongly correlated with the total luminosity of the galaxy, $L$,
in a relation of the form $L\propto V^\alpha$. The power $\alpha$
is around 3-4,
 and depends on the wavelength band at which $L$ is measured.
 The close agreement between this TF relation and the prediction
 of MOND is encouraging; but,
 to test MOND more precisely on this count, one would have
 to bridge properly the mass-asymptotic-velocity MOND relation
 with the commonly presented luminosity-bulk-velocity TF relation.
 One should use the luminosity in a band
where it is a good representative of the stellar mass, take into
account not only the stellar mass, as represented by the
luminosity, but also the contribution of gas to the mass, and use
the asymptotic velocity, as opposed to other measures of the
rotational velocity. It has emerged recently (see \cite{ver} and
reference therein) that if one does all this one indeed obtains a
tight and accurate relation of the form predicted by MOND.

\par
But, by far, the most clear-cut test of MOND is provided by
disc-galaxy rotation curves, simply because the astronomical
observations, and their interpretation, are the most complete and
best understood, if still not perfect. What we typically need to
know of a galaxy in order to apply this test has been discussed by
the various authors who conducted the test; for example,
\cite{bbs}, \cite{sanver}, \cite{sanrot}, \cite{mcdeb}. On the
whole, these tests speak cogently for MOND. These test involves
fitting the observed rotation curve of a galaxy by that predicted
by MOND. Such fits involve one free parameter per galaxy, which is
the assumed conversion factor from luminosity to mass in stars,
the so-called mass-to-light ratio. In fact, however, this
parameter is not totally free. It is constrained to an extent by
what theoretical understanding of galaxy composition tell us.
\cite{sanver} who have conducted a MOND rotation-curve analysis of
a sample of disc galaxies in the Ursa Major cluster, have compared
their deduced MOND best-fit $M/L$ values with theoretical results
from stellar-population synthesis. They found a very good
agreement. This shows that, to some extent, the MOND rotation
curves might be looked at as definite prediction of MOND, which
use theoretical $M/L$ values, and not as fits involving one free
parameter.

\par
Regarding galactic systems other than galaxies, the comparison of
the systematics of the observed mass discrepancy with the
expectations from MOND are shown in Figure 2 in \cite{mildark98}
based on analyzes referenced there. The agreement is uniform, with
one exception: The cores of rich x-ray clusters of galaxies show a
considerable mass discrepancy, while, according to MOND there
shouldn't be any, because the accelerations there are only of the
order of \a0, and not much smaller. (Application of MOND to the
clusters at large, say within a few megaparsecs of the center,
does predict correctly the mass discrepancy.) The resolution, by
MOND, will have to be that these cores harbor large quantities of
still undetected baryonic matter, perhaps in the form of dim
stars, perhaps as warm gas. The environment, and history, of these
cores is so unlike others that this would not be surprising.
\par
In order to appreciate the message that the phenomenological
success of MOND carries, we should note the following. According
to MOND, the acceleration constant \a0 appears in many independent
roles in the phenomenology of the mass discrepancy. For example,
in galaxies that have high central accelerations, the mass
discrepancy appears only beyond a certain radius; according to
MOND , the acceleration at this radius should always be \a0. \a0
also appears as the boundary acceleration between so called
high-surface-brightness galaxies (=high acceleration galaxies)
which do not show a mass discrepancy near the center, and
low-surface-brightness galaxies, where the discrepancy prevails
everywhere. \a0 appears in the relation between the asymptotic
rotational velocity of a galaxy and its total mass, and in the
mass-velocity relation for all sub-\a0 systems, etc., etc..
\par
These roles that \a0 plays are independent in the sense that in
the framework of the dark matter paradigm, it is easy to envisage
baryon-plus-dark-matter galactic systems that evince any of these
appearances of \a0 without showing the others. In other words, in
the dark matter paradigm one role of \a0 in the phenomenology does
not follow from the others.
\par
This is similar, for example, to the appearances of the Planck
constant in different quantum phenomena: in the black-body
spectrum, in the photoelectric effect, in the hydrogen spectrum,
in superconductivity, etc.. Phenomenologically, these roles of the
same constant seem totally unrelated. The only unifying frame is a
theory: non-relativistic quantum mechanics, in this case. MOND is,
likewise, a theory that in one fell swoop unifies all the above
appearances of \a0 in the phenomenology of galaxy dynamics.
\par
And finally, let me point out a possibly very significant
coincidence: The value of the acceleration constant \a0 that fits
all the data discussed above is about $10^{-8}cm~s^{-2}$. This
value of \a0 is of the order of some acceleration constants of
cosmological significance. It is of the same order as
 $a_{ex}\equiv cH_0$, where $H_0$ is the Hubble constant; and, it is
 also of the order of $a_{cc}\equiv c(\Lambda/3)^{1/2}$,
 where $\Lambda$ is the emerging value of the cosmological
 constant (or "dark energy"). So, for example, a body accelerating
 at \a0 from rest will approach the speed of light in the life
time of the universe.
\par
Because the cosmological state of the universe changes, such a
connection, if it is a lasting one, may imply that galaxy
evolution does not occur in isolation, affected only by nearby
objects, but is, in fact, responding constantly to changes in the
state of the universe at large. For example, if the connection of
$a_0$ with the Hubble constant always holds, the changing of the
Hubble constant would imply that $a_0$ must change over cosmic
times, and with it the appearance of galactic systems, whose
dynamics $a_0$ controls. If, on the other hand, \a0 is a
reflection of a true cosmological constant, then is might be a
veritable constant.

\section{MOND as an effective theory}
\par
But, on the more fundamental side, the above proximity may hint at
a deep connection between cosmology and local dynamics in systems
that are very small on cosmological scales. Either cosmology
somehow enters and affects local laws of physics, such as the law
of inertia or the law of gravity, or a common agent affects both
cosmology and local physics so as to leave on them the same
imprint. This would mean that MOND--and perhaps more cherished
notions, such as inertia--is a derived concept, or an effective
theory as we would say nowadays. An observed relation between
seemingly unrelated constants appearing in a theory (in our case,
\a0, the speed of light, and the radius of the horizon) may
indicate that it is only an approximation of a theory at a deeper
stratum, in which some of the constants do not really have any
special role.
 A parable will help clarify the point:
In experiments and observations confined to the vicinity of the
earth surface, there appears a constant: the free-fall
acceleration, $g$. If, for some reason, we were restricted to such
an ant world (for example because the earth is ever clothed in a
thick layer of clouds) unaware of planetary motions, universal
gravity, etc., we would have looked on $g$ as a true constant of
nature. We would also notice a mysterious relation between this
acceleration and two other important constants: the escape speed
$c_e$ (objects thrown with a higher velocity never return) and the
radius of the earth $R_\oplus$. This relation:
$g=c_e^2/2R_\oplus$, is practically the same as that between
$a_0$, the speed of light, and the Hubble radius, in MOND. But, we
do see beyond the earth's surface, and we do know about universal
gravity, which tells us that the "constants" $g$ and $c_e$
actually derive from the mass and radius of the earth (hence the
relation between the three). They are useful parameters when
describing near-earth-surface phenomena, but quite useless in most
other circumstances. In a similar vein, $a_0$ might turn out to be
a derived constant, perhaps variable on cosmic time scales,
perhaps even of no significance beyond the non-relativistic
regime, where MOND has been applied so far. Its connection with
the speed of light and the radius of the universe will, hopefully,
follow naturally in the underlying theory that still eludes us.
\par
Many instances of such effective theories are known. Even General
Relativity is now thought to be an effective, low-energy
approximation of a "higher" theory (e.g. a string-inspired
theory); an idea that has been anticipated by Sakharov's "induced
gravity" idea.

\section{Interpretations}
\par
Equations(\ref{basic})(\ref{mond})  have the form of a
modification of the law of inertia, but since they are algebraic
relations between the MOND and Newtonian accelerations they can
simply be inverted to read $a=F/m=a_Nf(a_N/a_0)$, which seems to
leave the second law intact, while modifying the Newtonian
gravitational force $ma_N$ to the MOND value $ma$. Because
gravitation is the sole force that governs galactic dynamics--the
only corner where the mass discrepancy has been clearly
observed--existing phenomenology does not distinguish well between
the interpretations of MOND as modified gravity, and modified
inertia. Although there are matter of principle differences
between the two interpretations (see below) they pertain to
observations that are not yet available. For now we must then
investigate both options.
\par
But what exactly is meant by modifying gravity, or modifying
inertia? When dealing with pure gravity the distinction is not
always clear. For example, the Brans-Dicke theory may be viewed as
either. But when other interactions are involved, the distinction
is clear. Obviously, modified inertia will enter the dynamics of
systems even when gravity is negligible, unlike the case for
modified gravity. Formally, the distinction might be made as
follows. In a theory governed by an action principle we
distinguish three part in the action: The pure gravitational part
(for example, the Einstein-Hilbert action in GR), the free action
of the matter degrees of freedom (in GR it also encapsules their
interaction with gravity), and the action of interactions between
matter degrees of freedom. By "modifying gravity" I mean modifying
the pure-gravity action; by "modifying inertia" I mean modifying
the kinetic ("free") matter actions.
\par
 To understand this definition consider that
 inertia is what endows the
motion of physical objects (particles, fields, large bodies, etc.)
with energy and momentum--a currency in the physical world. Motion
itself is only of a descriptive value; inertia puts a cost on it.
For each kind of object it tells us how much energy and momentum
we have to invest, or take away, to change its state of motion by
so much. This information is encapsuled in the kinetic action.
\par
For example, take the non-relativistic action for a system of
particles interacting through gravity.

\beq S=S_{\phi}+S_k+S_{in}=-(8\pi G)^{-1}\int d^3r~(\grad\phi)^2
+\sum_i(1/2)m_i\int dt~v_i^2-\int d^3r~\rho(\vr)\phi(\vr),
\label{acpar}\eeq
 where $\rho(\vr)=\sum_i
m_i\delta(\vr-\vr_i)$. (In GR, $S_k$ and $S_{in}$ are lumped
together into the particle kinetic action.)
\par
Here, modifying gravity would mean modifying $S_{\phi}$, while
modifying inertia would entail changing $S_k$.

\section{MOND as modified gravity}
\par
An implementation of MOND as a non-relativistic modified gravity
was discussed by \cite{bm}, who replaced the standard Poisson
action $S_{\phi}$ in eq.(\ref{acpar})  by an action of the form

\beq S_{\phi}=-(8\pi G)^{-1} a_0^2\int
d^3r~F[(\grad\phi)^2/a_0^2].\label{act} \eeq This gives, upon
variation on $\phi$, the equation

 \beq\div[\mu(|\grad\phi|)\grad\phi]=4\pi G\rho(\vr),
 \label{mondpoiss} \eeq
where $\mu(x)\equiv dF(y)/dy\vert_{y=x^2}$. This theory, since it
is derived from an action that has all the usual symmetries,
satisfies all the standard conservation laws. Its various
implications have been discussed in \cite{bm}, \cite{milsol},
\cite{milconformal}, and others.
\par
One important point to note is that this theory gives the desired
center-of-mass motion of composite systems: Stars, star clusters,
etc. moving in a galaxy with a low center-of-mass acceleration are
made of constituents whose internal accelerations are much higher
than \a0. If we look at individual constituents we see bodies
whose total accelerations are high and so whose overall motion is
very nearly Newtonian. Yet, their motion should somehow combine to
give a MOND motion for the center of mass. This is satisfied in
the above theory as shown in \cite{bm}. (A similar situation
exists in GR: imagine a system made of very tightly bound black
holes moving in the weak field of a galaxy, say. While the motions
of the individual components is highly relativistic, governed by a
non-linear theory, we know that these motions combine to give a
simple Newtonian motion for the center of mass.)
\par
This field equation, generically, requires numerical solution, but
it is straightforward to solve in cases of high symmetry
(spherical, cylindrical, or planar symmetry), where the
application of the Gauss law to eq.(\ref{mondpoiss}) gives the
exact algebraic relation between the MOND ($\vg=-\grad\phi$) and
Newtonian ($\vg_N=-\grad\phi_N$) acceleration fields:

\beq \mu(g/a_0)\vg=\vg_N, \label{algebraic} \eeq which is
identical to the heuristic MOND relation we started with. Note
that in general, for configurations of lower symmetry,  this
algebraic relation does not hold (and, in general, $\vg$ and
$\vg_N$ are not even parallel).
\par
It is worth pointing out that in such a modified-gravity theory,
the deep-MOND limit corresponds to a theory that is conformally
invariant, as discussed in \cite{milconformal}. Whether this has
some fundamental bearings is not clear, but it does make MOND
unique, and enables one to derive useful analytic results, such as
an expression for the two-body force, and a virial relation,
despite the obstacle of nonlinearity.
\par
 There is a large number of
physical phenomena that are governed by an equation like
eq.(\ref{mondpoiss}), each with its own form of the function
$\mu(x)$, as detailed in \cite{milconformal}, or \cite{milnonlin}.
I would like to concentrate here on one, in particular. It is well
known that a stationary, potential flow is described by the
Poisson equation: If the velocity field $\vu(\vr)$ is derived from
a potential, $\vu=\grad\phi$, then the continuity equation, which
here determines the flow, reads $\div\grad\phi=s(\vr)/\varrho_0$,
where $s(\vr)$ is the source density, and $\varrho_0$ is the
(constant) density of the fluid. When the fluid is compressible,
but still irrotational, and barotropic [i.e. has an equation of
state of the form $p=p(\rho)$] the stationary flow is described by
the nonlinear Poisson equation. The Euler equation reduces to
Bernoulli's law

\beq h(\varrho)=-u^2/2+const., \label{bern} \eeq where
$dh/d\rho\equiv \rho^{-1}dp/d\rho$. This tell us that $\varrho$ is
a function of $u=|\grad\phi|$. Substituting this in the continuity
equation gives

\beq \div[\varrho(|\grad\phi|)\grad\phi]=s(\vr), \label{flow} \eeq
which has the same form as eq.(\ref{mondpoiss}) if we identify
$\varrho$ as $\mu$, and the source density $s$ with the normalized
gravitational mass density $4\pi G\rho$ . Note, however, that from
the Bernoulli law, $d\varrho/d|u|= -\varrho|u|/c^2$, where
$c^2=dp/d\varrho$ is the formal squared speed of sound. Thus, in
the case of MOND, where we have that $\mu$ is an increasing
function of its argument, the model fluid has to have a negative
compressibility $c^2<0$.  A cosmological-constant equation of
state, $p=-c^2\varrho$, with $c$ the speed of light gives
$\varrho(u)=\varrho_0 exp(u^2/2c^2)$, which is not what we need
for MOND. The deep-MOND
 limit, $\mu(u)\approx u/a_0$, corresponds to
 $p=-(a_0^2/3)\varrho^3$. To get the Newtonian limit at large values
 of $u$ the equation of state has to become incompressible
 at some finite density $\varrho_0$, so that eq.(\ref{flow}) goes
 to the Poisson equation.
\par
 The gravitational force is then the pressure+drage force on
sources. For a small (test) static source $s$, at a position where
the fluid speed is $\vu$, the source imparts momentum to the flow
at a rate $s\vu$, and so is subject to a force $-s\vu$. The force
between sources of the same sign is attractive, as befits gravity.

 \par
 Note that in such a picture the fluid density itself $\varrho$
 does not contribute to the sources of the potential equation, so
 it does not, itself, gravitate. Also note that, because
$\rho=p=0$ for $\vu=0$, the fluid behaves as if it has no
 existence without the sources (masses) that induce velocities in
 it.
 Obviously, this picture is anything but directly applicable as
 an explanation of Newtonian gravity. For example, it is not clear
 how to obtain the barotropic equation of state that is needed to
 reproduce MOND. In particular, how does the infinite
 compressibility appear at a finite critical density, and what is the
 meaning of this density? Is this due to some phase transition?
 What happens at densities higher than this critical density? are
 they accessible at all? Also, there seem to be a drag force on
 moving sources. In the context of a time-dependent configuration
 the above equation of state is problematic.

\section{MOND as modified inertia}
\par
Most people seem to prefer modifying gravity to modifying inertia;
perhaps because the latter seems to be less drastic; perhaps
because it is a game that has been much played before. I
personally feel, without concrete evidence, that there is more
potential in modified inertia as the basis for MOND.
\par
 Remember first that Newtonian inertia has not been immune to changes.
 A familiar
modification of Newtonian inertia, which is taken to be "nature
given", is that brought about by Special Relativity. The
single-particle kinetic action in eq.(\ref{acpar}) is replaced by
$-mc^2\int~dt~[1-(v/c)^2]^{1/2}$, which gives an equation of
motion \beq \vec
F=md(\gamma\vv)/dt=m\gamma[\va+\gamma^2\vv(\vv\cdot\va)/c^2],
\label{lorentz} \eeq
 where $\gamma$ is the Lorentz factor.
\par
And, physics is replete with instances of modified, acquired, or
effective inertia. Electrons and holes in solids can sometimes be
described as having a greatly modified mass tensor. Mass
renormalization and the Higgs mechanism, modify particle masses
and/or endow them with mass: an effective, approximate description
that encapsules the effects of interactions of the particles, with
vacuum fields in the former instance, and with the Higgs field in
the latter. The effects of a fluid on a body embedded in it may
sometimes be described as a contribution to to the mass tensor of
the body, because its motion induces motion in the fluid which
carries energy and momentum. So, modified inertia might also well
lie in the basis of MOND.

\par
As a first stage of looking for Mondified inertia it might behoove
us to study non relativistic modifications of inertia that
incorporate the basic principle of MOND. We seek to modify the
particle kinetic action $S_k$ in eq.(\ref{acpar}) into an action
of the form $S_k[\vr(t),a_0]$, which is a functional of the
particle trajectory $\vr(t)$ and depends also on one constant,
\a0. It should satisfy the following asymptotic requirements: In
the formal limit $a_0\rightarrow 0$--corresponding to all
acceleration measures in the system being much larger than the
actual value of \a0 (this is similar to obtaining the classical
limit of quantum mechanics by taking the formal limit
$\hbar\rightarrow 0$)--it should go into the standard Newtonian
action. If we want to retain the MOND phenomenology, according to
which in the deep MOND limit $G$ and \a0 appear only through their
product $Ga_0$, then, in the limit $a_0\rightarrow \infty$,
$S_k\propto a_0^{-1}$. This can be seen by rescaling $\phi$ into
$\phi/G$ in eq.(\ref{acpar}) (and deviding the action by $G$).
\par
The theory should also satisfy the  more subtle requirement of the
correct center-of-mass motion discuss in the previous section.
\par
General properties of such theories are discussed in detail in
\cite{milinertia}. Here I summarize, very succinctly, some of the
main conclusions.
\par
If the particle free action enjoys the usual symmetries:
translational, rotational, and Galilei invariance, than to satisfy
the two limits in \a0 it must be non-local. This means that the
action cannot be written as $\int~L~dt$, where $L$ is a function
of a finite number of derivatives of $\vr(t)$. This might look
like a disadvantage, but, in fact, it is a blessing. A local
action for MOND would have had to be a higher-derivative theory,
and, as such, it would have suffered from the several severe
problems that beset such theories. A non-local theory need not
suffer from these. Indeed, I have discussed examples that are free
of these problems. A non-local action is also a more natural
candidate for an effective theory.
\par
While nonlocal theories tend to be rather unwieldy, they do lend
themselves to a straightforward treatment of the important issue
of rotation curves. This is done via a virial relation that
physical, bound trajectories can be shown to satisfy:

\beq 2S_k[\vr(t),a_0]-a_0{\partial S_k\over\partial a_0}=
<\vr\cdot\grad\phi>, \label{virial} \eeq where $\phi$ is the
(unmodified) potential in which the particle is moving, $<>$ marks
the time average over the trajectory, and $S_k$ is the value of
the action calculated for the particular trajectory ($S_k$ is
normalized to have dimensions of velocity square). In the
Newtonian case this reduces to the usual virial relation. Applying
this relation to circular orbits in an axi-symmetric potential,
and noting that, on dimensional grounds, on such orbits with
radius $r$ and velocity $v$ we must have
$S_k(r,v,a_0)=v^2\mu(v^2/ra_0)$, we end up with the expression for
the velocity curve

\beq (v^2/r)\mu(v^2/ra_0)=d\phi/dr. \label{roca} \eeq Thus the
algebraic relation that was first used in MOND as a naive
application of eq.(\ref{mond}), and which all existing
rotation-curve analyzes use, is exact in modified-inertia MOND. In
modified gravity this expression is a only good approximation.
\par
Another important difference between the two interpretations is
worth noting.
 Unlike (non-relativistic) modified gravity, where the
 gravitational field is modified, but in it all bodies at the same
 position undergo the same acceleration, in modified inertia the
 acceleration depends not only on position, but also on the
 trajectory. In the case of SR the acceleration depends on the
 velocity as well, but in more general theories it might depend on
 other properties of the orbit. There is
 still a generalized momentum whose rate of change {\it is} a
 function of position only ($m\gamma\vv$ in SR) but this rate is
 not the acceleration. This larger freedom in modified inertia
 comes about
 because we implement the modification via a
 modification of the action as a functional of the trajectory;
 namely,
 a function of an infinite number of variables; so,
 different trajectories might suffer different modifications.
  In modifying gravity we modify one function of the
three coordinates (the gravitational potential).
 This is an obvious point, but is worth making because in
 interpreting  data we equate observed accelerations with the
 gravitational field. While this is still true in modified gravity
 it is not so in modified inertia.
 \par
 We can exemplify this point by considering the claimed anomaly in
 the motions of the Pioneer 10 and 11 spacecraft. Analysis of
 their motion have shown an unexplained effect
 (see \cite{anderson}) that
 can be interpreted as being due to an unexplained constant
 acceleration towards the sun of about $7\cdot 10^{-8} cm~s^{-2}$,
 of the order of \a0. This might well be due to some systematic
 error, and not to new physics. This suspicion is strengthened by
 the fact that an addition of a constant acceleration of the above
 magnitude to the solar gravitational field is inconsistent with
 the observed planetary motions (e.g. it gives a much too large
 rate of planetary perihelion precession).
\par
 MOND could naturally
 explain such an anomalous acceleration: We are dealing here with
 the strongly Newtonian limit of MOND, for which we would have to
 know the behavior of the extrapolating function $\mu(x)$ at
 $x>>>1$, where $\mu\approx 1$.
  We cannot learn about this from galaxy dynamics, so we
 just parameterize $\mu$ in this region: $\mu\approx 1-\xi x^{-n}$.
 (This is not the most general form; e.g. $\mu$ may approach 1
 non analytically in $x^{-1}$, for example as $1-exp(-\xi x)$.)
 Be that as it may, if $n=1$ we get just the desired effect in
 MOND: the acceleration in the field of the sun becomes
 $M_\odot G r^{-2}+\xi a_0$ in the sun's direction. As
 I said above, in a modified gravity interpretation this would
 conflict with the observed planetary motions; but, in the
 modified-inertia approach it is not necessarily so. It may well
 be that the modification enters the Pioneers motion, which
 corresponds to unbound, hyperbolic motions, and the motion of
 bound, and quasi-circular trajectories in a different way. For example, the
 effective $\mu$ functions that correspond to these two motions
 might have different asymptotic powers $n$.

\section{vacuum effects and MOND inertia}
\par
Because MOND revolves around acceleration, which is so much in the
heart of inertia, one is directed, with the above imagery in mind,
to consider that inertia itself, not just MOND, is a derived
concept reflecting the interactions of bodies with some agent in
the background. The idea, which is as old as Newton's second law,
is the basic premise of the Mach's principle. The great sense that
this idea makes has lead many to attempt its implementation.
 The agent responsible for inertia had been taken to be
the totality of matter in the Universe.
\par
Arguably, an even better candidate for the inertia-producing
agent, which I have been considering since the early 1990s, in the
hope of understanding MOND's origin, is the vacuum. The vacuum is
known to be implicated in producing or modifying inertia; for
example, through mass renormalization effects, and through its
contribution to the free Maxwell action in the form of the
Euler-Heisenberg action  \cite{euler}. Another type of vacuum
contributions to inertia have been discussed by \cite{ja}. But, it
remains moot whether the vacuum can be fully responsible for
inertia.
\par
 The vacuum is thought to be Lorentz invariant, and so indifferent to
motion with constant speed. But acceleration is another matter. As
shown by Unruh in the 1970s, an accelerated body is alive to its
acceleration with respect to the vacuum, since it finds itself
immersed in a telltale radiation, a transmogrification of the
vacuum that reflects his accelerated motion.  For an observer on a
constant-acceleration ($a$) trajectory this radiation is thermal,
with $T=\alpha a $, where $\alpha\equiv \hbar/2\pi kc$. The effect
has been also calculated approximately for highly relativistic
circular motions; the spectrum is then not exactly thermal. In
general, it is expected that the effect is non-local; i.e.,
depends on the full trajectory.

\par
 Unruh's
result shows that the vacuum can serve as an inertial frame. But
this is only the first step. The remaining big question is how
exactly the vacuum might endow bodies with inertia. At any rate,
what we want is the full MOND law of inertia, with the transition
occurring at accelerations of order $a_0$ that is related to
cosmology. We then have to examine the vacuum in the context of
cosmology. How it affects, and is being affected by, cosmology.
One possible way in which cosmology might enter is through the
Gibbons-Hawking effect, whereby even inertial observers in an
expanding universe find themselves embedded in a palpable
radiation field that is an incarnation of the vacuum. The problem
has been solved for de Sitter Universe, which is characterized by
a single constant: the cosmological constant, $\Lambda$, which is
also the square of the (time independent) Hubble constant. In this
case the spectrum is also thermal with a temperature $T=\alpha c
(\Lambda/3)^{1/2}$.
\par
In the context of MOND it is interesting to know what sort of
radiation an observer sees, who is accelerated in a non-trivial
universe: If the Unruh temperature is related to inertia, then it
might be revealing to learn how this temperature is affected by
cosmology. This can be gotten for the case of a
constant-acceleration observer in a de Sitter Universe. For this
case the radiation is thermal with a temperature
$T=\alpha(a^2+c^2\Lambda/3)^{1/2}$ \cite{deser}. Inertia, which is
related to the departure of the trajectory from that of an
inertial observer, who in de Sitter space sees a temperature
$\alpha c(\Lambda/3)^{1/2}$, might be proportional to the
temperature difference \beq \Delta
T=\alpha[(a^2+c^2\Lambda/3)^{1/2}-c(\Lambda/3)^{1/2}],
\label{det}\eeq and this behaves exactly as MOND inertia should:
it is proportional to $a$ for $a>>a_0\equiv 2c(\Lambda/3)^{1/2}$,
and to $a^2/a_0$ for $a<<a_0$; and, we reproduce the connection of
\a0 with cosmology. Of course, in the modified-inertia paradigm
this would reflect on a "linear", constant-acceleration motion,
while circular trajectories will probably behave differently. But
the emergence of an expression {\it a-la} MOND in this connection
with the vacuum is very interesting.

\section{Relativistic theories}
\par
We still want a relativistic extension of MOND. Such a theory is
needed for conceptual completion of the MOND idea. But, it is
doubly needed because we already have observed relativistic
phenomena that show mass discrepancies, and we must ascertain that
there too the culprit is not dark matter but modified dynamics.
\par
 Because of the
near values of \a0 and the Hubble acceleration, there are no local
black holes that are in the MOND regime. The only system that is
strongly general relativistic and in the MOND regime is the
Universe at large. This, however means that we would need a
relativistic extension of MOND to describe cosmology. In fact, as
I have indicated, MOND itself may derive from cosmology, so it is
possible that the two problems will have to be tackled together as
parts and parcels of a unified concept. And, because the
cosmological expansion is strongly coupled with the process of
structure formation this too will have to await a modified
relativistic dynamics for its treatment.
\par
Several relativistic theories incorporating the MOND principle
have been discussed in the literature, but none is wholly
satisfactory (see, e.g. \cite{bm}, \cite{pcg}, \cite{stratified},
and references therein).
\par
There have also been attempts to supplement MOND with extra
assumptions that will enable the study of structure formation, so
as to get some glimpse of structure formation in MOND. For these
see \cite{comments}, \cite{sanstructure}, and \cite{nusser}.
\par
Gravitational light deflection, and lensing, is another phenomenon
that requires modified relativistic dynamics. It is tempting to
take as a first approximation the deflection law of post-Newtonian
General Relativity with a potential that is the non-relativistic
MOND potential (see e.g. analyzes by \cite{qin}, and
\cite{mortlock} based on this assumption). This, however, is in no
way guaranteed. In GR this is only a post-Newtonian approximation,
and perhaps it would turn out to be a post-Newtonian approximation
of MOND (i.e. an approximation of MOND in the almost Newtonian,
$a>>a_0$ regime). But, there is no reason to assume that it is
correct in the deep-MOND regime. Even in the framework of this
assumption one needs to exercise care. For example, the thin-lens
hypothesis, by which it is a good approximation to assume that all
deflecting masses are projected on the same plane perpendicular to
the line of sight, breaks down in MOND. For example, $n$ masses,
$M$, arranged along the line of sight (at inter-mass distances
larger that the impact parameter) bend light by a factor $n^{1/2}$
more than a single mass $nM$.
\par
Also note that we may expect surprises in mondified inertia where
we cannot even speak of the modified, MOND potential, as alluded
to above.

\end{document}